\documentclass[11pt,twoside]{article}  
\usepackage{adassconf}

\begin{document}

\paperID{P01.10}

\title{Towards a Large Field of View Archive for the European VLBI
  Network}
\titlemark{Towards a Large Field of View Archive for the EVN}
\author{Huib Jan van Langevelde, Friso Olnon, Harro Verkouter, 
Bauke Kramer, Mike Garrett, Arpad Szomoru, Steve
Parsley}

\affil{Joint Institute for VLBI in Europe, Postbus 2, 7990 AA
Dwingeloo, The Netherlands}

\author{Albert Bos}
\affil{ASTRON, Postbus 2, 7990 AA Dwingeloo, The Netherlands}

\contact{Huib Jan van Langevelde}
\email{langevelde@jive.nl}

\paindex{van Langevelde, H.J.}
\aindex{Olnon F.M.}
\aindex{Verkouter H.}
\aindex{Kramer B.}
\aindex{Garrett M.A.}
\aindex{Szomoru A.}
\aindex{Parsley S.M.}
\aindex{Bos A.}

\authormark{van Langevelde, Olnon, Verkouter, Kramer et al.}


\begin{abstract}
Traditionally VLBI observations focus on a small patch of sky and image
typically a few 100 mas around a bright source, which is often used to
self-calibrate the data. High spectral and time resolution is needed to
image a larger area, in principle up to the primary beam of the
individual telescopes. The EVN MkIV data processor at JIVE is being
upgraded to make such high resolution data its standard product.  From the
archive of high resolution data it will be possible to image many
sources in each field of view around the original targets.
\end{abstract}

\section{Introduction}

The input data rate of a VLBI processor sets the total bandwidth that
can be processed from each telescope. The output data rate determines
how fine the correlation product can be sampled in frequency and time.
The frequency resolution is important for various spectral line
applications, but it also limits the field of view (FoV) that can be imaged
before bandwidth smearing sets in. The temporal sampling also puts
constraints on the FoV through time smearing, which scales
with baseline length and distance from the field centre.  The spectral
resolution of a VLBI correlator is usually determined by the computing
power built into its hardware, e.g.\ the number of available lags per
baseline. The product over all available telescope pairs of spectral
resolution and short visibility integration time combines into a large
total output rate. There is generally a limit on the datarate at which
this can be flushed out to a standard computing environment and saved
on disk.

Both the spectral capabilities and the output bandwidth of the EVN data
processor at JIVE are being upgraded.  The first by introducing
recirculation and the latter by the PCInt project. In this paper we
discuss the scientific motivation and the future data handling of this
system.

\section{Scientific Justification} 

A VLBI dataset, if properly calibrated, has flat phase response, both
in time and frequency, for the target position. Positional offsets from
the phase centre introduce increasingly steep phase slopes. As long as
these are properly sampled, the structure of
sources away from the centre can be derived without time or bandwidth
smearing. Both effects scale with baseline length (and are therefore
particularly severe for VLBI); time smearing also scales with
frequency (Wrobel 1995). In a traditional continuum VLBI
experiment the integration time may be as long as 4s and typically $8
\times 8$ MHz bands are sampled, each with 16 spectral points. The
resulting limits on the FoV can be seen in
Table~\ref{P01.10_t1}.  Although the original recordings of a single
VLBI experiment hold information over the whole field, typically
only $10^3$ out of $10^8$ beams are imaged.

\begin{deluxetable}{lrrrrrr}
\scriptsize
\tablecaption{The field of view set by integration time and spectral
resolution .\label{P01.10_t1}}
\tablehead{
Application & \colhead{$N_{\rm sp}$} & \colhead{$t_{\rm int}$}  &   
\colhead{Output} & 
\colhead{FoV$_{\rm t}$} & \colhead{FoV$_{\rm bw}$} &
\colhead{$V_{\rm 12hr}$} \nl
 &  & \colhead{[s]} & \colhead{[MB/s]} & \colhead{[$'$]} & 
\colhead{[$'$]} & \colhead{[GB]} \nl
}
\startdata
Traditional      &   128  & 4.000 & 0.02  & 0.70  & 0.82 &   1 \nl
Operational max  &  1024  & 0.500 & 1.50  & 5.57  & 6.59 &  63 \nl
Phase 0          &  2048  & 0.250 & 6.00  &11.14  &13.19 & 253 \nl
Full system      &  4096  & 0.031 &96.00  &89.11  &26.36 &4050 \nl
\enddata
\end{deluxetable}

There are several astronomical applications that require larger fields
of view. Galactic masers may extend over rather large areas, especially
in star formation complexes. Gravitational lenses are another
case where VLBI sources may extend over a large FoV.
Moreover, studies of the faint radio source population may be done 
more efficiently when a large instantaneous field of view is
available.  A long integration at the full recording
bandwidth can then be employed to study many weak sources simultaneously.
As an example we consider the high resolution observations of radio
sources in the Hubble Deep Field (HDF). It contains many radio-sources
at low flux level over a large (by VLBI standards) field. Long
integrations with the most sensitive telescopes are required to
investigate their nature with VLBI.  The Effelsberg beam encompasses an
area much larger than the HDF and so in principle a
single observation can be used to study the nature of each source with
VLBI. This technique was explored with the EVN at 1.6 GHz and the VLBA
correlator by Garrett et al. (2001). Even at this moderate resolution
the FoV barely covers the HDF. VLBI detections for 3
sources were made at 150-350 $\mu$Jy.  Such studies will benefit
greatly from the upgrades ongoing in the EVN, both in recording and
correlator capacity.

\section{Dataflow}

The EVN MkIV data processor correlates inputs from 16 stations
simultaneously (Schilizzi et al.\ 2001). Each telescope input can
handle up to 1 Gbit/s, from Mk4 tape or Mk5 disk playback. Its
computing power is based on 32 boards, each equipped with 32 custom
made chips, each producing 256 complex lags. This yields sufficient
spectral capabilities to attribute 512 spectral channels to every
baseline between 16 telescopes. 

\begin{figure}
\epsscale{.60}
\plotone{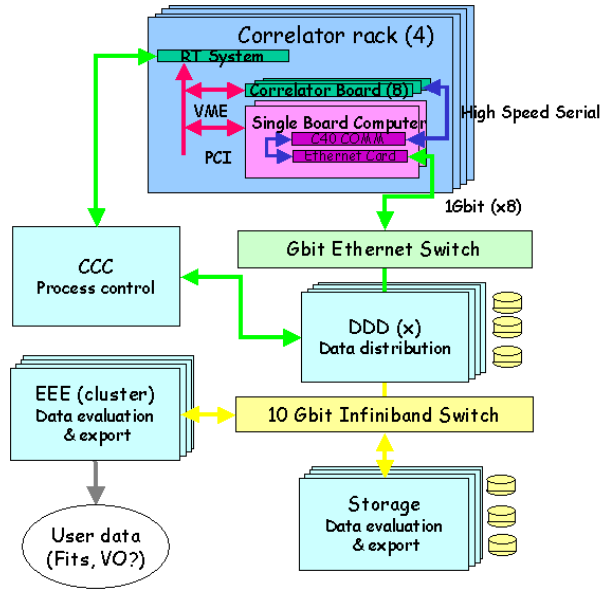}
\caption{Outline of the data flow for the PCInt project}\label{P01.10_f1}
\end{figure}

In its original configuration groups of 8 boards were controlled by an
HP-RT system, which flushed the data out on 4 parallel 10Mb/s Ethernet
lines.  The first improvement to this system has recently been
implemented by upgrading the system with 8 Single Board Computers
(SBC), which handle the data from the HP-RT processors and have 100
Mb/s Ethernet to flush the data (Phase 0). In the final PCInt
configuration each rack will have two Single Board Computers running
Linux, which read the data from DSP powered serial ports. A total of
$8\times 1$ GB/s Ethernet connections are then available to flush out
the data. The software is set up in such a way that the data can be
handled in parallel by a set of workstations that write the data to an
array of disks with fast access. This is necessary to overcome another
possible bottleneck: disk access. The architecture can be seen in
Figure~\ref{P01.10_f1}.

In table~\ref{P01.10_t1} the Field of View (FoV) limits set by
the bandwidth sampling (bw) and time smearing (t) in various stages of
the project are shown. The calculations have been performed for a
rather modest VLBI recording at 18cm on 8 EVN antennas with
a total bandwidth of 64 MHz (2 bit sampled). The primary beam of a 25m
telescope measures 27'. The requirements become even more severe at
higher spatial resolution (global baselines, or higher frequency).

In the original system the data is first collected in raw correlator
format. Downstream the data is transformed into an aips++
MeasurementSet.  After local calibration and data quality control,
FITS files are written. An archive of user products and diagnostic
plots is on-line through a \htmladdnormallinkfoot{web
  interface}{http://www.jive.nl/archive/scripts/listarch.php}. 
Initially the data is password protected while the PI has the
proprietary right.

With PCInt the data streams will increase, but similar operations on the
data are still necessary. Data quality evaluation and internal
calibration must be performed promptly. We are investigating
whether this inspection can be performed without a physical copy of the
data to aips++ internal format.  While the output data will be at full
resolution, the PI may receive the data at a coarser resolution, one that
is optimal for the scientific goal of his study. In a similar way the
interface to the archive will allow users to make a selection and
create a dataset at a lower resolution, possibly by averaging for a
new target position.  This operation will be ported to a parallel
processing environment, in order to make such products available in an
almost interactive manner.

Through pipeline processing, JIVE is already providing preliminary
calibration for every dataset. In order to accommodate this service for
the data archive product, the calibration data must be closely
integrated. Special care will be required for the calibration of data
with a new field centre. In addition, the projected output data rate
yields datasets of such large sizes that the astronomer will probably
require tailor-made software and dedicated hardware to process them.
Solutions for these issues are being investigated and will involve
parallel computing. It is possible that wide field of view VLBI images
will be made as an integral part of the data processor product.  Such a
solution would fit in with the Virtual Observatory paradigm.

After the successful completion of the so-called phase 0 project (Nov
2003), the European VLBI Network welcomes proposals that use the full
correlator at 0.25s read-out (6MB/s).  The data flow and software for
the next phase are being tested and more capacity upgrades are expected
in 2004. Then there will be a focus on the hardware and software to
process all data at the maximum resolution and compute the data product
from the archive using parallel processing. This effort has recently
required funding from the EU for 2004-2007.

\end{document}